\documentclass[10pt]{article}
\usepackage{amsfonts}
\newcommand {\sla}[1]{ #1 \!\!\!/}
\usepackage{epsfig,cite}
\begin{document}
\begin{center}
{\Large\bf $\Delta$ contribution in $e^+ + e^- \rightarrow p + \bar{p}$ at small $s$}\\
\vspace*{1cm}
Hai Qing Zhou $^1 \footnote{E-mail: zhouhq@mail.ihep.ac.cn}$,
Dian Yong Chen $^2$ and Yu Bing Dong$^{2,3}$\\
\vspace{0.3cm} %
{\ $^1$ Department of Physics, Southeast University,
 Nanjing,\ 211189,\ P. R. China}\\
{\ $^2$ Institute of High Energy Physics,
 Chinese Academy of Science,\ Beijing,\ 100049,\ P. R. China}\\
{\ $^3$ Theoretical Physics Center for Science Facilities, CAS,
Beijing 100049, China}
\vspace*{1cm}
\end{center}
\begin{abstract}

Two-photon annihilate contributions in the process $e^+ + e^-
\rightarrow p + \bar{p}$ including $N$ and $\Delta$ intermediate are
discussed in a simple hadronic model. The corrections to the
unpolarized cross section and polarized observables $P_x,P_z$ are
presented. The results show the two-photon annihilate correction to
unpolarized cross section is small and its angle dependence becomes
weak at small $s$ after considering the $N$ and $\Delta(1232)$
contributions simultaneously, while the correction to $P_z$ is
enhanced.
\end{abstract}

\section{Introduction}

The Two-Photon-Exchange(TPE) effect has attracted many interests
after its success in explaining the un-consistent measurements of
$R=\mu_pG_E/G_M$ from $ep \rightarrow ep$ by Rosenbluth technique
and polarized methods\cite{Jone00,Blunden03,Chen04,Blunden05}. It is
found that the TPE corrections play an important role in extracting
the proton's form factors due to its explicit angle dependence.
Later some other processes\cite{Rekola99,Gakh05,Tomaso08} are
suggested to measure the TPE like effects. The $e^+ + e^-
\rightarrow p + \bar{p}$ is one of such processes and the two-photon
annihilate corrections in this process have been discussed  by
\cite{DianYong08} where only the $N$ intermediate was included. The
estimate by \cite{DianYong08} showed the two-photon annihilate
corrections are about a few percent in the magnitude but strongly
depend on the hadron production angle. On another hand, the
calculation in \cite{Kondratyuk05} showed the $\Delta(1232)$
intermediates also unneglectable in the TPE corrections in the
simple hadronic model \cite{Blunden03,Blunden05,Kondratyuk05}. These
researches prompt us to extent the estimate of the two-photon
annihilate corrections in \cite{DianYong08} to include $\Delta$
intermediate state. In this work, we present such results.

\section{Two-Photon Annihilate Corrections including $N$ and $\Delta(1232)$ as Intermediate State}
Considering the process $e^+(k_2) + e^-(k_1) \rightarrow p(p_2) +
\bar{p}(p_1)$, the Born diagram is showed as Fig.1. The differential
cross section for this process at the tree level can be written
as\cite{Cabibbo61}
\begin{eqnarray}
(\frac{d\sigma}{d\Omega})_{CM}=\frac{\alpha^2\sqrt{1-4M_N^2/q^2}}{4q^2}(|G_M|^2(1+cos^2\theta)+\frac{1}{\tau}|G_E|^2sin^2\theta).
\end{eqnarray}
where $q=k_1+k_2, \tau=q^2/4M_N^2>1$ and $\theta$ is the angle
between the momentum of finial antiproton and initial electron in
the center of mass frame. The Sachs form factors have been used as
\begin{eqnarray}
G_M(q^2)=F_1(q^2)+F_2(q^2), G_E(q^2)=F_1(q^2)+\tau F_2(q^2).
\end{eqnarray}

In principle, the form factors at certain $s=q^2$ can be extracted
from the measurement of the unpolarized differential cross section
at different angle. To extract the form factors more precisely, the
radiative corrections should be considered. Among the one loop
radiative corrections, the box and crossed box diagrams play special
role due to their strong angle dependence. This leads us restrict
our discussions on the two-photon annihilate correction firstly.
\begin{figure}[t]
\centerline{\epsfxsize 3.0 truein\epsfbox{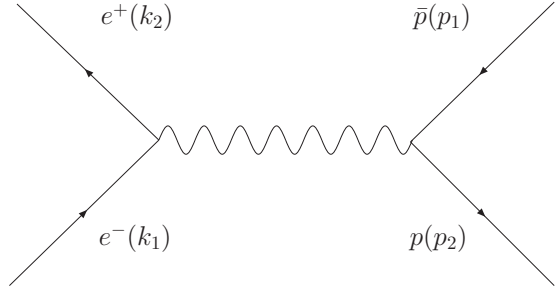} } \caption{ One
photon annihilating diagram for $e^+ + e^- \rightarrow p + \bar{p}$.
}
\end{figure}
\begin{figure}[t]
\centerline{ \epsfxsize 2.8 truein\epsfbox{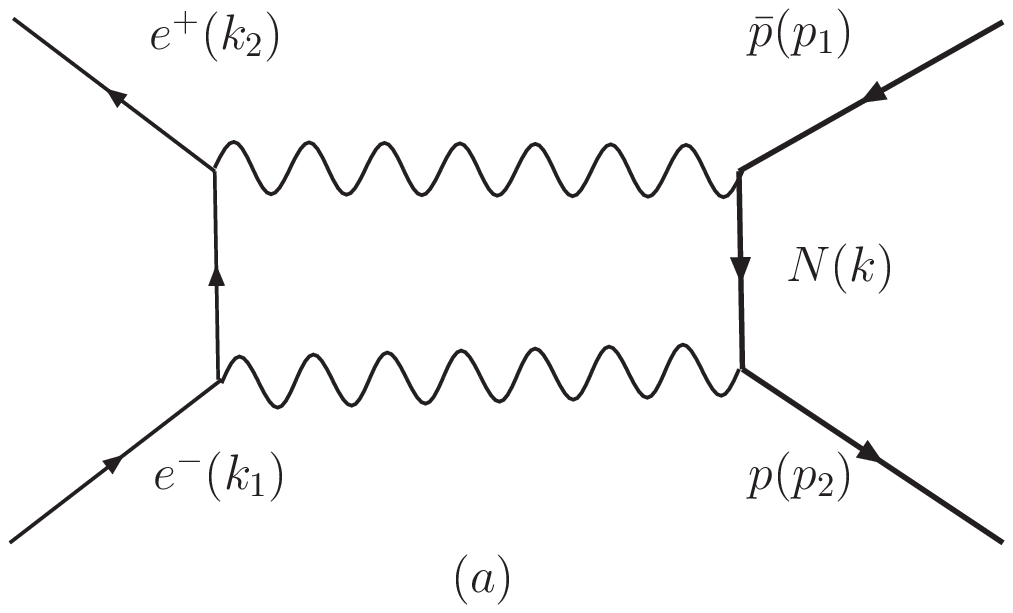}\epsfxsize 2.8
truein\epsfbox{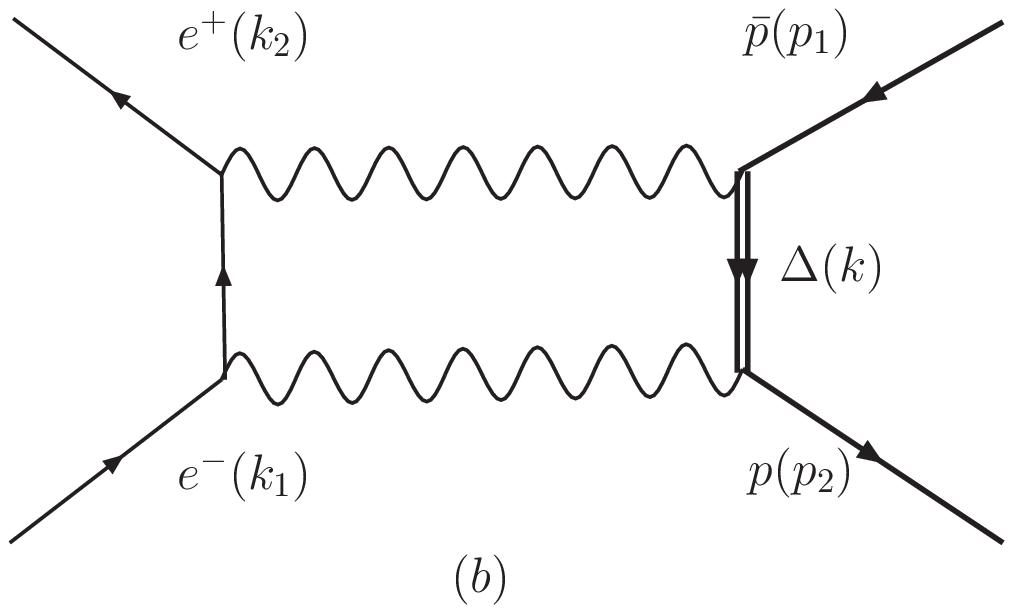} } \caption{Two-photon annihilating
diagrams (a) with $N$ as intermediate state,(b) with $\Delta(1232)$
as intermediate state. Corresponding cross-box diagrams are
implied.}
\end{figure}

Using the simple hadronic model developed in
\cite{Blunden03,Blunden05,Kondratyuk05} and including $N$ and
$\Delta$ as the intermediate state like Fig.2, the unpolarized cross
section can be written as
\begin{eqnarray}
d\sigma=d\sigma_0(1+\delta_N^{2 \gamma}+\delta_\Delta^{2
\gamma})\propto \sum\limits_{helicity}
{|\mathcal{M}_0+\mathcal{M}^{2\gamma}_N+\mathcal{M}_\Delta^{2\gamma}|^2},
\end{eqnarray}
where $\mathcal{M}_0$ is the contribution of  one-photon annihilate
diagram and $\mathcal{M}^{2\gamma}_{N,\Delta}$ denote the
contribution from two-photon annihilate diagrams with $N$ and
$\Delta$ as intermediate state. The corrections to the unpolarized
cross section can defined as
\begin{eqnarray}
\delta_{N,\Delta}^{2 \gamma}=
\frac{\sum\limits_{helicity}{2Re\{{\mathcal{M}_{N,\Delta}^{2\gamma}
\mathcal{M}_0^\dagger}\}}}{\sum\limits_{helicity}{|\mathcal{M}_0|^2}}.
\label{delta}%
\end{eqnarray}
The corrections from $N$ have been discussed in \cite{DianYong08}.
To discuss the correction from $\Delta$, we take the following
matrix elements as \cite{Kondratyuk05,Keitaro08}
\begin{eqnarray}
&&\langle N(p_2)|J^{em}_\mu|\Delta (k)\rangle =\frac{
-F_{\Delta}(q_1^2)}{M_{N}^{2}}\overline{u}(p_2)
[g_{1}(g^{\alpha}_{\mu}\sla{k}\sla{q}_1
-k_{\mu}\gamma^{\alpha}\sla{q}_1
-\gamma_{\mu}\gamma^{\alpha}k\cdot q_1+\gamma_{\mu}\sla{k}q_1^{\alpha})\nonumber \\
&&+g_{2}(k_{\mu}q_1^{\alpha}-k\cdot q_1 g^{\alpha}_{\mu})
+g_{3}/M_{N} (q_1^2(k_{\mu}\gamma^{\alpha}-g^{\alpha}_{\mu}\sla{k})
+q_{1\mu}(q_1^{\alpha}\sla{k}-\gamma^{\alpha}k\cdot
q_1))]\gamma_{5}T_3u_{\alpha}^{\Delta}(k),\nonumber \\
\nonumber \\
&&\langle \Delta (k) \overline{N}(p_1)|J^{em}_\nu|0\rangle =\frac{
-F_{\Delta}(q_2^2)}{M_{N}^{2}}\overline{u}_{\beta}^{\Delta}(k)T_3^+\gamma_{5}
[g_{1}(g^{\beta}_{\nu}\sla{q}_2\sla{k}
-k_{\nu}\sla{q}_2\gamma^{\beta}
-\gamma^{\beta}\gamma_{\nu}k\cdot q_2+\sla{k}\gamma_{\nu}q_2^{\beta})\nonumber \\
&&+g_{2}(k_{\nu}q_2^{\beta}-k\cdot q_2 g^{\beta}_{\nu}) -g_{3}/M_{N}
(q_2^2(k_{\nu}\gamma^{\beta}-g^{\beta}_{\nu}\sla{k})
+q_{2\nu}(q_2^{\beta}\sla{k}-\gamma^{\beta}k\cdot q_2))]v(p_1),
\label{GamaNDelta}
\end{eqnarray}
where $q_1=p_2-k,~q_2=k+p_1$ and $T_3$ is the third component of the
$N\rightarrow \Delta$ isospin transition operator and is
$-\sqrt{2/3}$ here. The effective vertexes of $\gamma N\Delta$ are
defined as
$\overline{u}(p_2)\Gamma_\mu^\alpha(\gamma\Delta\rightarrow N)
u_{\alpha}^{\Delta}(k)=-ie\langle N(p_2)|J^{em}_\mu|\Delta
(k)\rangle,~\overline{u}_{\beta}^{\Delta}(k)\Gamma_\nu^\beta(\gamma\rightarrow
\overline{N}\Delta) v(p_1)=-ie\langle \Delta (k)
\overline{N}(p_1)|J^{em}_\nu|0\rangle$. Both the two vertexes
satisfy the conditions $q_{1,2}^\mu\Gamma_\mu=0$ and
$k_\alpha\Gamma^\alpha=0$, the first condition ensure the gauge
invariance of the result and the second condition ensure to select
only the physical spin3/2 component \cite{Kondratyuk05}.

For the propagator of $\Delta$, the same form is employed as
\cite{Kondratyuk05}
\begin{eqnarray}
&&S_{\alpha\beta}^{\Delta}(k)=\frac{-i(\sla{k}+M_\Delta)}{k^2-M_\Delta^2+i\epsilon}P_{\alpha\beta}^{3/2}(k),\nonumber \\
&&P_{\alpha\beta}^{3/2}(k)=g_{\alpha\beta}-\gamma_{\alpha}\gamma_{\beta}/3
-(\sla{k}\gamma_{\alpha}k_{\beta}+k_{\alpha}\gamma_{\beta}\sla{k})/3k^2.
\end{eqnarray}
Such propagator is different with the usual R.S one which read as
\begin{eqnarray}
S_{\alpha\beta}^{RS}(k)=\frac{\sla{k}+M_\Delta}{k^2-M_\Delta^2+i\epsilon}[-g_{\alpha\beta}+\frac{1}{3}\gamma_\alpha\gamma_\beta
+\frac{1}{3m}(\gamma_\alpha k_\beta-\gamma_\beta
k_\alpha)+\frac{2}{3m^2}k_\alpha k_\beta].
\end{eqnarray}
After using the properties of the vertexes, these two forms result
in the same amplitude.

By this effective interaction, the amplitude of box diagram Fig.2(b)
can be written as
\begin{eqnarray}
&&M^{(2b)}=-i\int\frac{d^4k}{(2\pi)^4}\overline{u}(k_2)(-ie\gamma_{\mu})\frac{i(\sla{p}_1+\sla{k}-\sla{k}_2+m_e)}
{(p_1+k-k_2)^2-m_e^2+i\varepsilon} (-ie\gamma_{\nu})
v(k_1)\frac{-i}{(p_1+k)^2+i\varepsilon}\frac{-i}{(p_2-k)^2+i\varepsilon}\nonumber \\
&&~~~~~~~~~~~~~\overline{u}(p_2)
\Gamma^{\mu\alpha}_{\gamma\Delta\rightarrow N}
\frac{-i(\sla{k}+M_{\Delta})}
{k^2-M_{\Delta}^2+i\varepsilon}P_{\alpha\beta}^{3/2}(k)\Gamma^{\beta\nu}_{\gamma\rightarrow\overline{N}
\Delta}v(p_1),
\end{eqnarray}
where  Feynamn gauge invariance has been used. Similarly one can get
the amplitude of crossed box diagram with $\Delta$ intermediate
state.

In the practical calculation, we take the form factor $F_\Delta$ in
the monopole form as $G_E$ in $N$ case\cite{DianYong08}
\begin{eqnarray}
F_\Delta(q^2)=G_E(q^2)=G_M/\mu
_p(q^2)=\frac{-\Lambda_1^2}{q^2-\Lambda_1^2},
\end{eqnarray}
the coupling parameters and cut-offs are the same as
\cite{DianYong08,Keitaro08}
\begin{eqnarray}
g_1=1.91,g_2=2.63,g_3=1.58,\Lambda_1=0.84GeV.
\end{eqnarray}

\section{Numerical Results and Discussion}
\begin{figure}[t]
\centerline{ \epsfxsize 2.8 truein\epsfbox{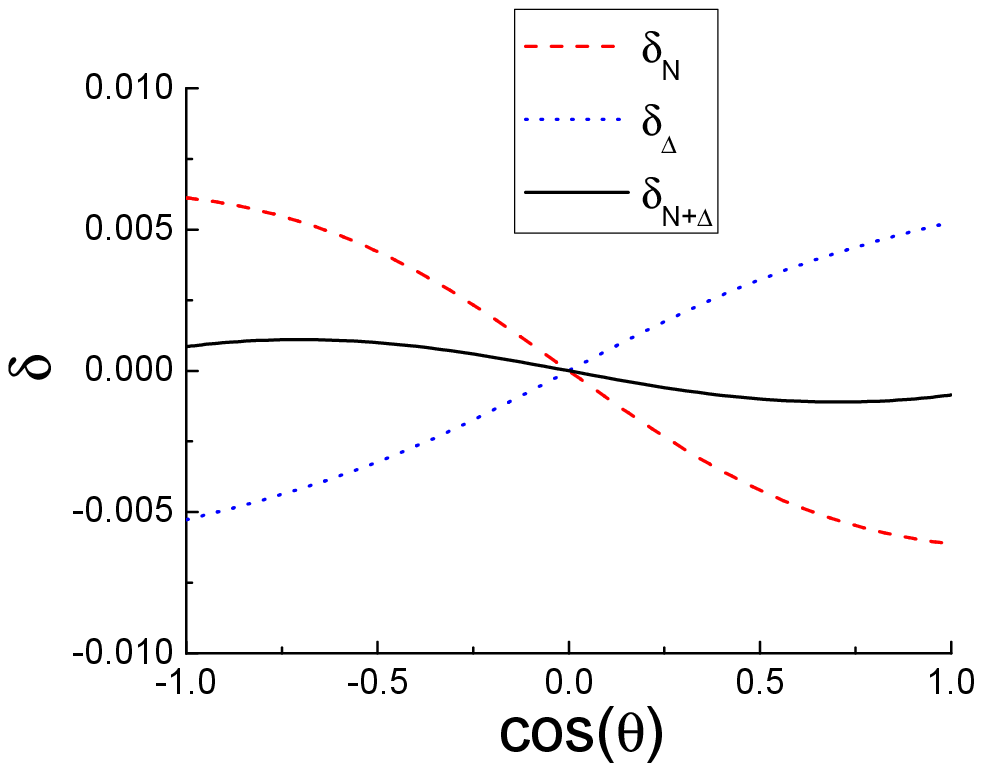}\epsfxsize 2.8
truein\epsfbox{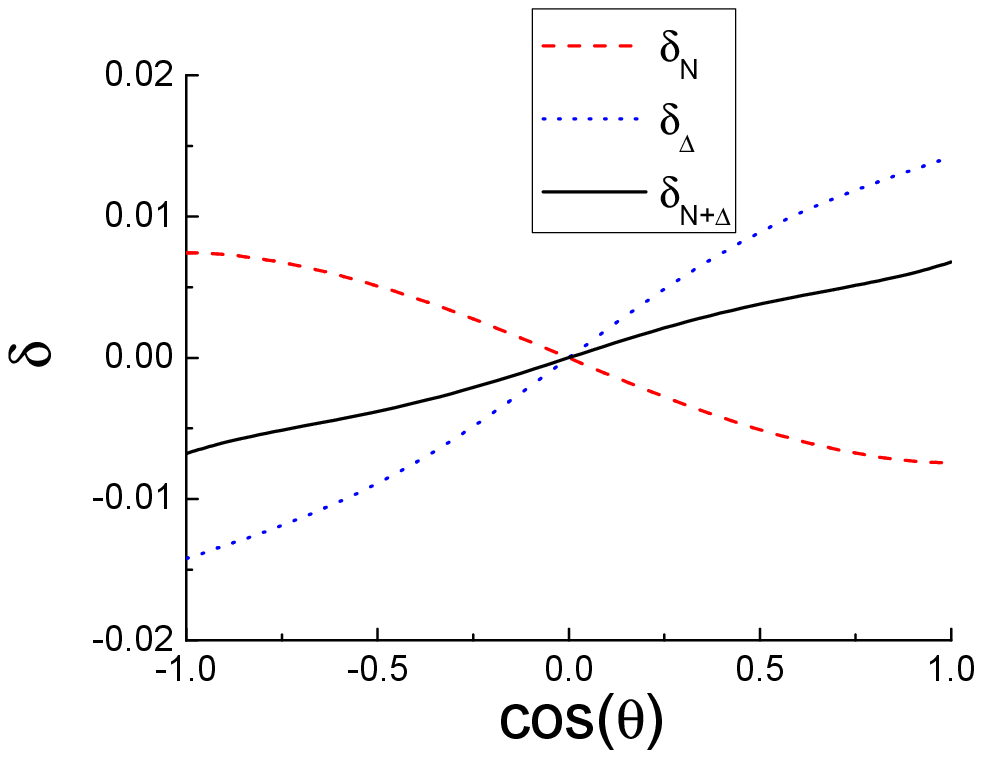} } \caption{Cosine $\theta$ dependence of
two-photon-annihilating corrections to unpolarized cross section.
The dashed and dotted lines denote to the correction from $N$ and
$\delta_\Delta$, respectively, and their sum is given by the solid
lines. The left result is for $s=4GeV^2$ and the right one for
$s=5GeV^2$}
\end{figure}
\begin{figure}[t]
\centerline{ \epsfxsize 2.8 truein\epsfbox{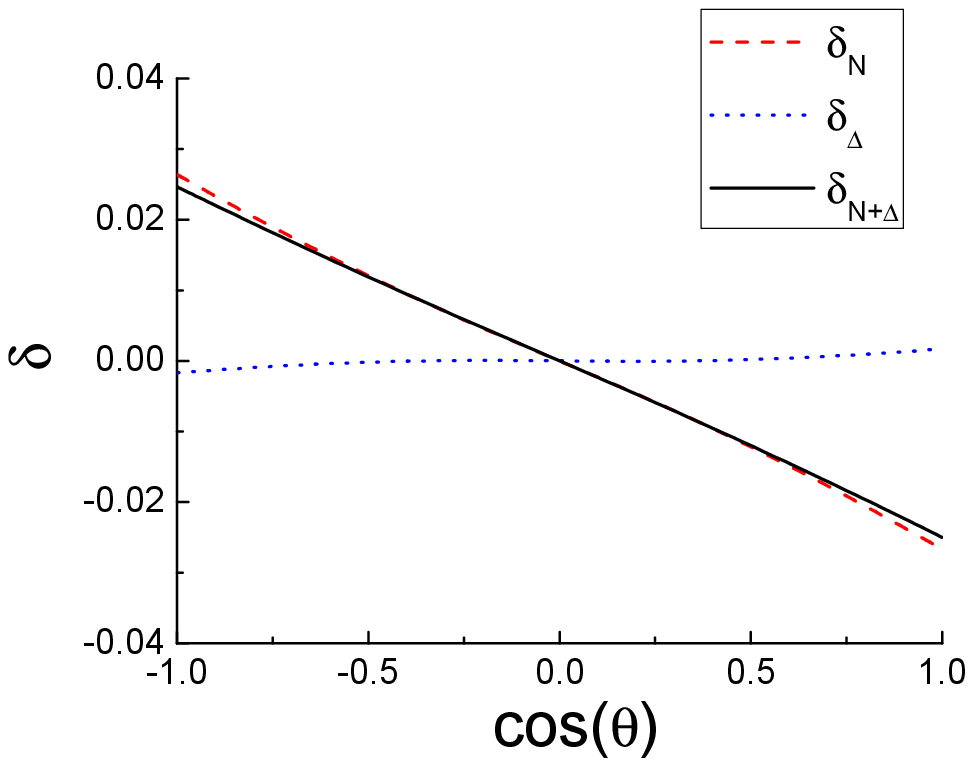}\epsfxsize 2.8
truein\epsfbox{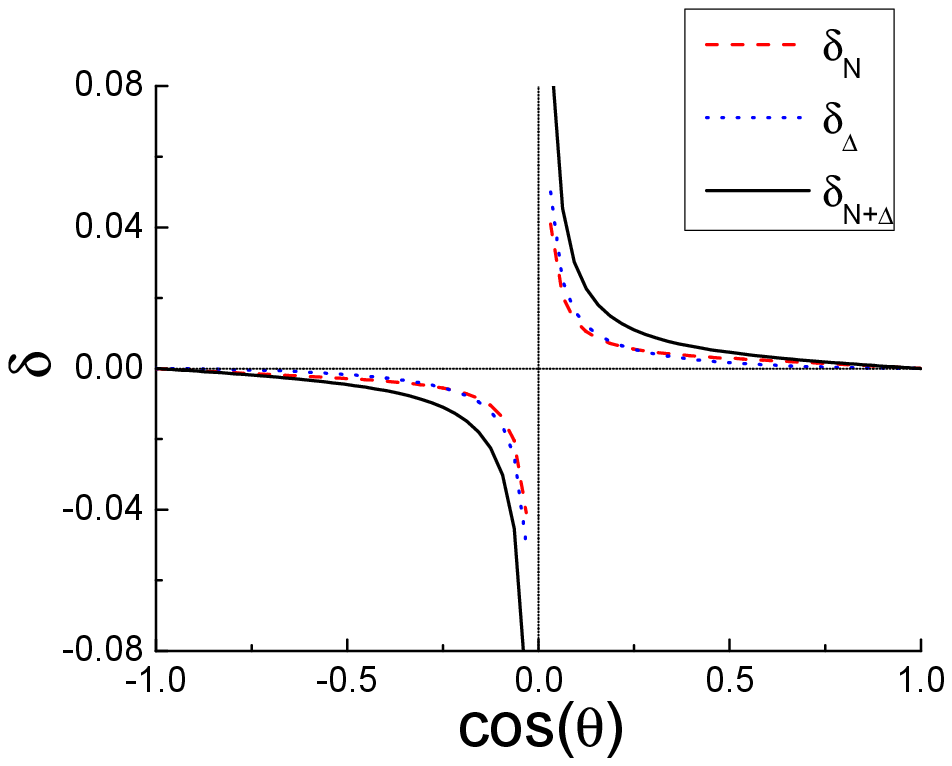} } \caption{Cosine $\theta$ dependence of
two-photon-annihilating corrections to $P_x$ and $P_z$. The dashed
and dotted lines denote to the correction from $N$ and
$\delta_\Delta$, respectively, and their sum is given by the solid
lines. The left result is for $P_x$ and the right one for $P_z$,
both with $s=4GeV^2$.}
\end{figure}
Using the above as input,  the two-photon annihilate corrections can
be calculated directly. We use the package FeynCalc \cite{FeynCalc}
and LoopTools \cite{LoopTools} to carry out the calculation. The IR
divergence in the  $N$ intermediate case is treated as
\cite{DianYong08} and there is no divergence in the $\Delta(1232)$
case. The numerical results for $\delta_{N,\Delta}^{2\gamma}$ are
showed in Fig.3. The similar calculation can be applied to the
polarized quantities $P_x$ and $P_z$ as \cite{0704.3375,DianYong08}
with the definitions
\begin{eqnarray}
\frac{d\sigma}{d\Omega}=\frac{d\sigma_{un}}{d\Omega}[1+P_y\xi_y+\lambda_eP_x\xi_x+\lambda_eP_z\xi_z].
\end{eqnarray}
The results of the corrections to $P_x$ and $P_z$ are presented in
Fig4. In our previous results\cite{DianYong08}, when discussing the
TPE corrections to polarized observables, only the contributions in
term $\frac{d\sigma}{d\Omega}$ are considered, while the corrections
in $\frac{d\sigma_{un}}{d\Omega}$ are neglected. Here the
calculations are improved to include both corrections.

As showed in Fig.3, the correction $\delta^{2\gamma}_\Delta$ is
found to be always opposite to the corrections $\delta^{2\gamma}_N$
in all the angle region. This behavior is similar to the  $ep$
scattering case \cite{Kondratyuk05}. Detailedly, at $s=4GeV^2$ the
absolute magnitude of $\delta^{2\gamma}_\Delta$ is so close to
$\delta^{2\gamma}_N$ which results in the large cancelation and
small total correction to unpolarized cross section. The small
$\delta_{N+\Delta}^{2\gamma}$ and its weak angle dependence suggest
the Rosenbluth method will work well in this region. This conclusion
is some different with the $ep$ scattering case where the
cancelation is much smaller and the total correction still strongly
depend on the scattering angle. At $s=5GeV^2$, the absolute
magnitude of $\delta^{2\gamma}_\Delta$ becomes larger than
$\delta^{2\gamma}_N$ which suggests the important roles played by
$\Delta(1232)$ intermediate state in the process of $e^+ + e^-
\rightarrow p + \bar{p}$.

For the polarized observables, Fig.4 shows the correction to $P_x$
from $\Delta$ is much smaller than $N$ and the correction to $P_z$
from $\Delta$ is close to $N$. The former property suggests
$\Delta(1232)$ gives no new correction than \cite{DianYong08} while
the latter property increases the two-photon annihilate corrections
to $P_z$ which enhances our previous suggestion that the nonzero
$P_z$ at $\theta=\pi/2$ may be a good place to measure the
two-photon exchange like effects directly.

\section{Acknowledgment}
This work is supported by the National Sciences Foundations of China
under Grant No.10747118, No.10805009, No. 10475088, and by CAS
Knowledge Innovation Project No. KC2-SW-N02.


\end{document}